\documentclass[
aps,
prl,
reprint,
groupedaddress,
showpacs,
]{revtex4-1}

\usepackage [utf8]{inputenc}

\usepackage{amssymb,amsmath}
\usepackage[
]{graphicx}
\usepackage{xcolor}
\usepackage{dsfont}

\usepackage[english]{babel}

\usepackage{siunitx}

\usepackage{pgfplots}

\usetikzlibrary{arrows,shapes,backgrounds,calc,
    positioning,
    intersections}
\newcommand{\ket}[1]{\left| #1 \right>} 

\newcommand{\unitb}{\hat{\mathbf{b}}}
\newcommand{\unitn}{\hat{\mathbf{n}}}

\newcommand{\unitu}{\hat{\mathbf{u}}}

\newcommand{\unitx}{\hat{\mathbf{x}}}
\newcommand{\unity}{\hat{\mathbf{y}}}
\newcommand{\unitz}{\hat{\mathbf{z}}}

\newcommand{\Bbold}{\mathbf{B}}
\newcommand{\Jbold}{\mathbf{J}}
\newcommand{\mbold}{\mathbf{m}}

\usepackage{layouts}
\usepackage{float}

\begin{document}

\title{
Enhanced magnetic sensitivity with non-gaussian quantum fluctuations
}

\author{Alexandre Evrard}
\author{Vasiliy Makhalov}
\author{Thomas Chalopin}
\author{Leonid A. Sidorenkov}
\altaffiliation{LNE-SYRTE, Observatoire de Paris, Universit\'e PSL, CNRS, Sorbonne Universit\'e, 61 Avenue de l'Observatoire, 75014 Paris, France}
\author{Jean Dalibard}
\author{Raphael Lopes}
\author{Sylvain Nascimbene}
\email{sylvain.nascimbene@lkb.ens.fr}

\affiliation{Laboratoire Kastler Brossel,  Coll\`ege de France, CNRS, ENS-PSL University, Sorbonne Universit\'e, 11 Place Marcelin Berthelot, 75005 Paris, France}

\date{\today}
\begin{abstract}
\noindent 
The precision of a quantum sensor can overcome its classical counterpart when its constituents are entangled. 
In gaussian squeezed states, quantum correlations lead to a reduction of the quantum projection noise below the shot noise limit. However, the most sensitive states involve complex non-gaussian quantum fluctuations, making the required measurement protocol challenging. Here we measure the sensitivity of non-classical states of the electronic spin $J=8$ of dysprosium atoms, created using light-induced non-linear spin coupling. Magnetic sublevel resolution enables us to reach the optimal sensitivity of non-gaussian (oversqueezed) states, well above the capability of squeezed states
 and  about half the Heisenberg limit. 
\end{abstract}

\maketitle

The measurement of a physical quantity is fundamentally limited in precision by the quantum nature of the measurement apparatus, via the Heisenberg uncertainty principle \cite{helstrom_quantum_1969-1,degen_quantum_2017-2}. Similarly to the mere averaging of $N$ independent measurements, a measurement device made of $N$ independent quantum probes allows reducing the measurement uncertainty by a factor $\sqrt{N}$ compared to a single realization, leading to the standard quantum limit of precision (SQL). Conversely, a set of correlated quantum probes may reach a better sensitivity \cite{caves_quantum-mechanical_1981-2,wineland_spin_1992-1}, ultimately up to the Heisenberg limit -- a measurement uncertainty reduced by a factor $N$ \cite{giovannetti_advances_2011-2}. However, reaching this precision limit with large-size quantum systems remains challenging, as it requires manipulating highly entangled quantum states, whose increased measurement sensitivity comes together with a higher fragility to environmental perturbations \cite{demkowicz-dobrzanski_chapter_2015}. 

A quantum sensitivity enhancement has been demonstrated in various experimental settings, including photonic systems \cite{holland_interferometric_1993,higgins_entanglement-free_2007,pan_multiphoton_2012}, trapped ions \cite{meyer_experimental_2001,
leibfried_creation_2005,roos_designer_2006,monz_14-qubit_2011-1,bohnet_quantum_2016-1},  Rydberg atoms \cite{facon_sensitive_2016-1}, thermal atomic gases  \cite{kuzmich_generation_2000,appel_mesoscopic_2009,leroux_implementation_2010,bohnet_reduced_2014-1,
hosten_measurement_2016,chalopin_quantum-enhanced_2018} or Bose-Einstein condensates \cite{sorensen_many-particle_2001-1,esteve_squeezing_2008-1,riedel_atom-chip-based_2010,jaskula_sub-poissonian_2010,bucker_twin-atom_2011,bookjans_strong_2011,lucke_twin_2011-2,hamley_spin-nematic_2012-1,strobel_fisher_2014-1,luo_deterministic_2017}. In squeezed quantum states described by gaussian statistics,  fluctuations of the mean response of the $N$ probes are reduced below the shot noise limit, thus increasing the measurement precision \cite{caves_quantum-mechanical_1981-2}. In the most common squeezing protocols the measurement uncertainty is decreased by a factor $N^{2/3}$ intermediate between the SQL and the Heisenberg limit \cite{kitagawa_squeezed_1993,pezze_quantum_2018-1}. The  precision can be further improved using  states with non-gaussian quantum fluctuations, characterized by high-order correlations between all probes  \cite{gessner_metrological_2018-1}. Quantum sensing with such non-gaussian states has been demonstrated in Refs.\;\cite{strobel_fisher_2014-1,bohnet_quantum_2016-1}; yet, the reported spectroscopic enhancement values remain limited, as reaching optimal sensitivity typically requires single-particle resolution \cite{zhang_collective_2012-1,hume_accurate_2013} or non-linear detection \cite{yurke_su2_1986,davis_approaching_2016-2,linnemann_quantum-enhanced_2016-1,nolan_optimal_2017-1}.  

\begin{figure}
\hspace*{-0.02\linewidth}\includegraphics[
trim={0 6mm 0 0.4cm},
draft=false,scale=0.89]{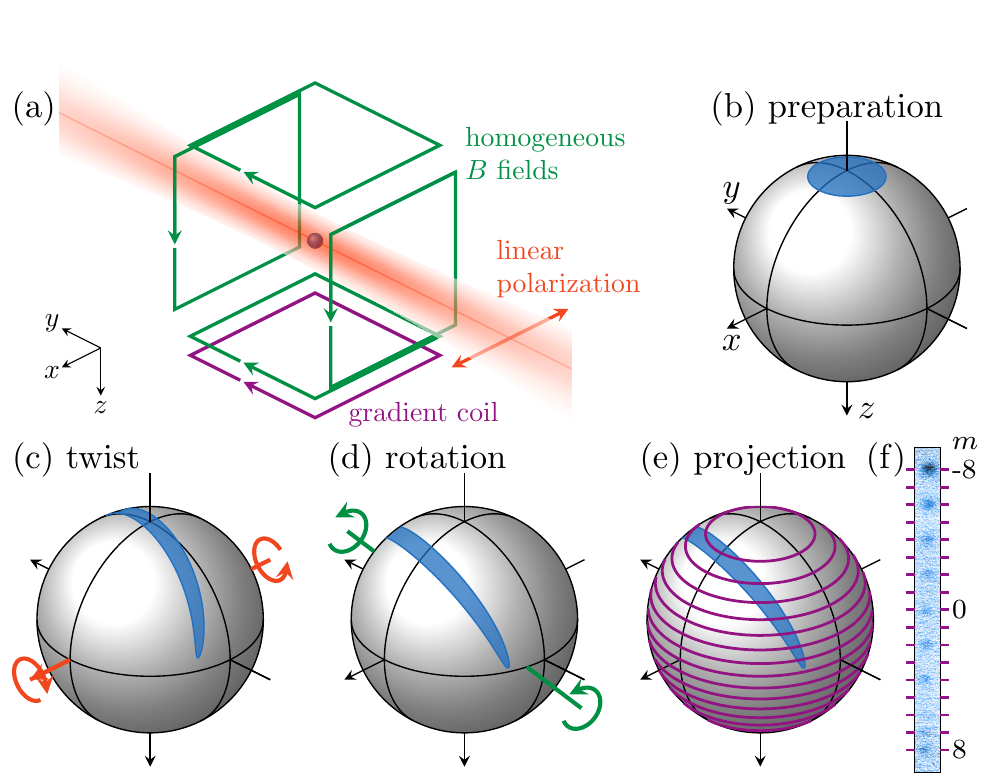}
\caption{
(a) Scheme of the experimental setup. Starting with a coherent state of the electronic spin of Dysprosium atoms aligned with the south pole (b), we induce non-linear dynamics using an off-resonant laser beam  (c). We then perform a spin rotation (d) followed by a projective measurement along $z$ using a magnetic field gradient (e). A typical absorption image is shown in (f).
\label{fig_scheme}}
\end{figure}

In this Letter, we use ultracold samples of atomic Dysprosium to study the magnetic-field sensitivity of gaussian and non-gaussian quantum spin states, encoded for each atom in its electronic spin of size $J=8$ -- equivalent to a set of precisely $N=2J=16$ elementary spin-1/2 particles \cite{landau_quantum_1958}. We use spin-dependent light shifts to induce non-linear dynamics described by the one-axis twisting Hamiltonian $\hat H=\hbar\chi \hat J_x^2$ \cite{kitagawa_squeezed_1993}.
These dynamics generate gaussian squeezed states at short times, before the stretching of spin distribution leads to non-gaussian `oversqueezed' states. 
Single magnetic sublevel resolution gives us access to the magnetic sensitivity hidden in non-gaussian quantum fluctuations, yielding a spectroscopic enhancement  of 8.6(6) compared to the SQL, consistent with the maximum sensitivity $J+\frac{1}{2}$ expected for oversqueezed states and about half the Heisenberg limit. We stress that our method is not based on correlations between different atoms but rather exploits the spin degree of freedom of individual atoms. A clear asset for our procedure robustness is the  absence of effective constituents number fluctuations $N=2J$. 

The experimental protocol is pictured in  Fig.\;\ref{fig_scheme}. We first prepare a gas of  $1.0(2)\times10^5$ atoms of $^{162}$Dy at a temperature $T=\SI{1.1(2)}{\micro\kelvin}$, using standard cooling techniques \cite{chalopin_anisotropic_2018}. The atoms are initially spin-polarized in the absolute ground state $\ket{m=-J}_z$, under a quantization field $\Bbold=B\unitz$, with $B=\SI{60.6(3)}{\milli G}$. We shine on the atoms an off-resonant laser beam  inducing spin-dependent light shifts thanks to the proximity to the narrow 626-nm optical transition (natural linewidth $\Gamma\simeq\SI{0.85}{\micro\second^{-1}}$). For a linear light polarization along $\unitx$, the light shift reduces (up to a constant) to a coupling $\hbar\chi \hat J_x^2$, where the rate $\chi$ is proportional to the light intensity (in the range $\chi\sim1-\SI{10}{\micro\second^{-1}}$) \cite{smith_continuous_2004-2,Note3}. Over the typical pulse duration, $t\sim\SI{100}{\nano\second}$, the Larmor rotation induced by the quantization magnetic field is $\sim\SI{3}{\degree}$ only, and we neglect it hereafter. We thus expect the dynamics to be well described solely by the one-axis twisting Hamiltonian (see Fig.\;\ref{fig_scheme}c).
After the non-linear spin dynamics, we apply time-dependent magnetic fields to rotate the spin along arbitrary directions (see Fig.\;\ref{fig_scheme}d). We finally perform a projective measurement along $z$ using a magnetic field gradient that spatially separates the $\ket{m}_z$ magnetic sublevels after a free expansion of \SI{2.45}{ms} (see Fig.\;\ref{fig_scheme}e,f). Combining rotation and projective measurement gives us access to the projection probabilities $\Pi_m(\unitn)$ ($-J\leq m\leq J$) along any direction $\unitn$ \cite{Note1}.

\begin{figure}
\hspace*{-0.03\linewidth}
\includegraphics[
draft=false,scale=0.86
]{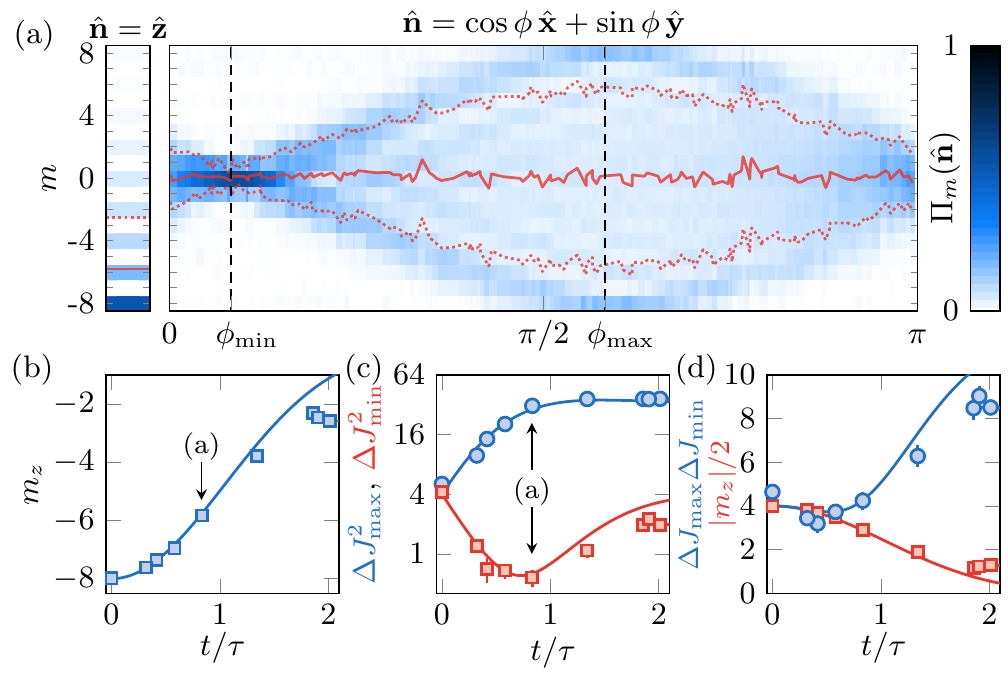}
\vspace{-7mm}
\caption{
(a) Projection probabilities $\Pi_m(\unitn)$ along $\unitn=\unitz$ and $\unitn\perp\unitz$, for  an interaction time $t=0.83(1)\tau$, with $\tau\!=\!(\sqrt{2J}\chi)^{-1}$. The solid (dotted) red line indicates the magnetization $m_{\unitn}$ (values of $m_{\unitn}\pm\Delta J_{\unitn}$).
(b) Magnetization $m_z$ as a function of the interaction time $t$. 
(c) Maximum and minimum spin projection variances $\Delta J_{\mathrm{max}}^2$ and $\Delta J_{\mathrm{min}}^2$ (blue dots and red squares, respectively).
(d) Comparison between the uncertainty product $\Delta J_{\mathrm{max}}\Delta J_{\mathrm{min}}$ and the half mean spin length $|m_z|/2$.
The solid lines in (b,c,d) correspond to the one-axis twisting model predictions. In all figures of this article error bars represent the 1-$\sigma$ statistical uncertainty determined using a bootstrap sampling method.
\label{fig_variances}}
\end{figure}
 
We first characterize the produced spin states by measuring their first and second spin moments. We expect from the symmetry of the one-axis twisting  Hamiltonian  that the mean spin $\mbold\equiv\langle\hat\Jbold\rangle$ remains oriented along $z$. An example of populations $\Pi_m(\unitz)$ is shown in Fig.\;\ref{fig_variances}a, from which we extract the magnetization $m_z$. We find that the magnetization decreases with time in absolute value as expected from the one-axis twisting model (see Fig.\;\ref{fig_variances}b). We also plot in Fig.\;\ref{fig_variances}a projection probabilities measured along directions $\unitn\perp\unitz$, from which we extract the minimum (maximum)  uncertainty $\Delta J_{\mathrm{min}}$  ($\Delta J_{\mathrm{max}}$), for a projection direction $\unitn_{\mathrm{min}}$ ($\unitn_{\mathrm{max}}$) of azimutal angle  $\phi_{\mathrm{min}}$ ($\phi_{\mathrm{max}}$, respectively). 

For $t=0$, the spin is polarized in $\ket{-J}_z$, corresponding to a coherent spin state. This state constitutes the best representation of a classical state magnetized along $-\unitz$, with zero magnetization along $x$ and $y$, and projection uncertainties $\Delta J_x/J=\Delta J_y/J=1/\sqrt{2J}$ taking the minimum value allowed for an isotropic distribution in the $xy$ plane \cite{arecchi_atomic_1972-2}. For this state, we find that for all directions $\unitn\perp\unitz$ the population distributions remain identical, and the projection variance $\Delta J_{\unitn}^2\!=4.3(2)$, as expected \cite{arecchi_atomic_1972-2}. For $t>0$,  we measure a squeezing of the minimum projection uncertainty down to $\Delta J_{\mathrm{min}}^2\!=0.6(1)$, i.e. about 7 times smaller than the coherent state value (see Fig.\;\ref{fig_variances}c).  The maximum spin quadrature $\Delta J_{\mathrm{max}}^2$ increases with $t$ up to a value $\simeq 37(1)$. This behavior is consistent with a semi-classical picture of  spin `diffusion' over the entire $yz$ meridian, leading to steady asymptotic values \mbox{$\Delta J_{\mathrm{min}}^2\! =\!\Delta J_{x}^2\!=\!J/2$} and \mbox{$\Delta J_{\mathrm{max}}^2\!=\!\Delta J_{y}^2\!=\!\Delta J_{z}^2\!=\!J(J+\frac{1}{2})/2\!=\!34$}. We find this dynamics to occur on the timescale of the diffusion time \mbox{$\tau\equiv(\sqrt{2J}\chi)^{-1}$} expected within the one-axis twisting model \cite{kitagawa_squeezed_1993}. We also use these measurements to quantify the gaussian character of quantum fluctuations, characterized by a saturation of the  Heisenberg uncertainty relation $\Delta J_{\mathrm{max}}\Delta J_{\mathrm{min}}\!\geq|m_z|/2$ \cite{holevo_gaussian_2011}. As shown in Fig.\,\ref{fig_variances}d, we find that this inequality is saturated for $t<0.5\tau$, while non-gaussian states occur for longer times.
 
\begin{figure}
\hspace*{-0.03\linewidth}
\includegraphics[
draft=false,scale=0.95
]{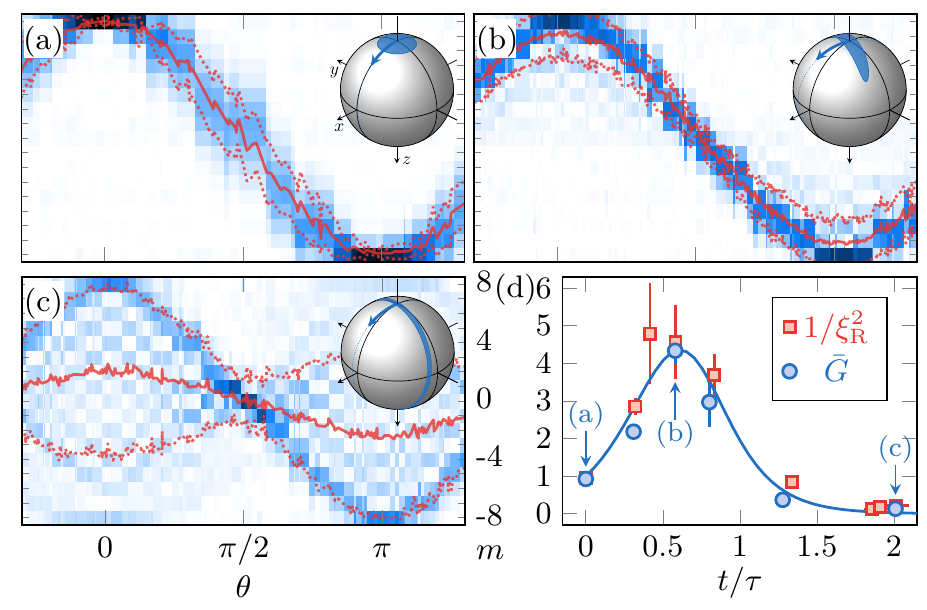}
\vspace{-7mm}
\caption{
(a,b,c) Evolution of the projection probabilities $\Pi_m$ upon a Larmor rotation of angle $\theta$ around the direction $\unitn_{\mathrm{max}}$ of maximum sensitivity, for a coherent state (a), a squeezed state (b, interaction time $t=0.58(2)\tau$) and an oversqueezed state (c, $t=2.01(1)\tau$). The solid (dotted) red line corresponds to the magnetization $m_z$ (values of $m_z\pm\Delta  J_z$) computed from the $\Pi_m$ values. 
(d) Usual metrological gain $\bar G$ and value of $1/\xi_{\mathrm{R}}^2$ deduced from the Fig.\;\ref{fig_variances}b,c data as a function of the interaction time $t$. The solid  line corresponds to  the one-axis twisting model prediction.
\label{fig_Ramsey}}
\end{figure}

We now discuss magnetic field sensing, i.e. the estimation of small rotation angles $\nu$ around an axis $\unitb$. In the most basic scheme, one estimates the angle $\nu$ from a measurement of the mean spin projection, giving access to the magnetization $m_z$ up to the projection noise $\Delta J_z$. The single-shot uncertainty on the estimation of  $\nu$ then reads $\Delta\nu=\Delta J_z/|dm_z/d\nu|$ \cite{helstrom_quantum_1969-1}.
For a set of $N=2J$ uncorrelated spins $\frac{1}{2}$, optimal sensitivity $\Delta\nu_{\mathrm{SQL}}=1/\sqrt{2J}$ is expected when all probes are aligned together, corresponding to a coherent spin state \cite{arecchi_atomic_1972-2}, and for a rotation axis $\unitb\perp\mbold$.  To check this behavior, we measure the precession of the coherent state $\ket{m=-J}_z$ around a direction $\unitb\perp\unitz$, parametrized by the angle $\theta$ 
 (see Fig.\,\ref{fig_Ramsey}a). We estimate the sensitivity of the state obtained after a rotation $\theta_0=\pi/2$ by evaluating the slope $dm_z/d\nu=-8.01(4)$ at the vicinity of $\theta_0$. We extract, at this angle, a value of $\Delta J_z^2=4.3(1)$, leading to $\Delta\nu=1.04(3)\,\Delta\nu_{\mathrm{SQL}}$, which validates our procedure.

We extend this measurement to the states produced after non-linear dynamics. We observe a decrease of the magnetization oscillation amplitude corresponding to the reduction of the  mean spin length (see Fig.\,\ref{fig_Ramsey}b,c). The best magnetic sensitivity is achieved for a rotation axis $\unitb$ coinciding with the direction $\unitn_{\mathrm{max}}$ of maximal spin projection variance $\Delta J_{\mathrm{max}}$, and around $\theta=\pi/2$. We quantify the increase of sensitivity with respect to the SQL by the metrological gain 
$
\bar G\equiv(\Delta\nu_{\mathrm{SQL}}/\Delta\nu)^2
$
 \cite{pezze_quantum_2014-1}.
 For durations \mbox{$0<t<\tau$} we observe a quantum enhancement  $\bar G>1$, with a maximum gain $\bar G=4.3(4)$ reached for \mbox{$t=0.58(2)\tau$}.  As shown in  Fig.\,\ref{fig_Ramsey}d, our data are in good agreement with the one-axis twisting model predictions \cite{kitagawa_squeezed_1993}. We expect the sensitivity to be related to the minimum spin projection variance $\Delta J_{\mathrm{min}}$, as $\bar G=1/\xi_{\mathrm{R}}^2$, where we introduce the so-called spin squeezing parameter $\xi_{\mathrm{R}}\equiv\sqrt{2J}\Delta J_{\mathrm{min}}/|m_z|$ \cite{wineland_spin_1992-1}. We verify this relation in  Fig.\,\ref{fig_Ramsey}d, where  the $\xi_{\mathrm{R}}$ values are computed from the measured $m_z$ and $\Delta J_{\mathrm{min}}$ data. For $t>\tau$, we observe that the gain $\bar G$  drops below unity, as expected from  the  mean spin length reduction.

\begin{figure}
\hspace*{-0.0\linewidth}\includegraphics[
draft=false,scale=0.9,trim={4mm 4mm 0 0.cm},
]{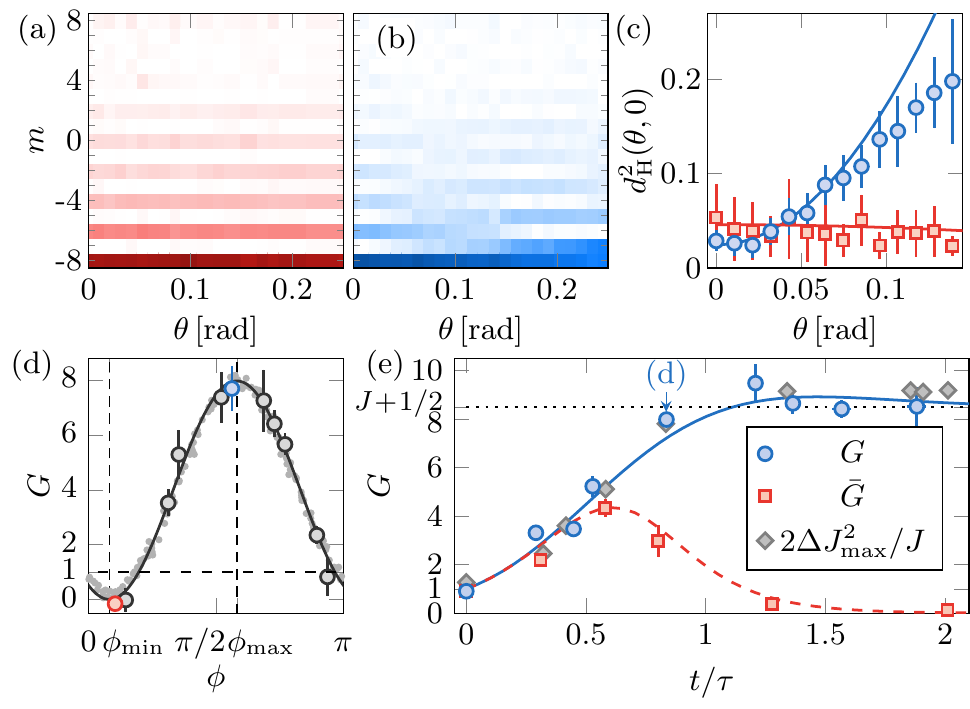}
\caption{
(a,b) Projection probabilities $\Pi_m$ measured for small rotation angles $\theta$ around $\unitb=\cos\phi\,\unitx+\sin\phi\,\unity$, with $\phi=0.10(2)\pi\simeq\phi_{\mathrm{min}}$ and $0.56(2)\pi\simeq\phi_{\mathrm{max}}$, respectively, for an interaction time $t=0.835(5)\tau$. Each probability is the average of 3 independent experiments.
(c) Hellinger distances $d_{\mathrm{H}}^2(\theta,0)$ deduced from (a,b), together with a quadratic fit of the small-$\theta$ data.
(d) Metrological gain $G$ deduced from the curvature of the Hellinger distance as a function of the azimutal angle $\phi$ (gray circles). The black line is a sine fit of the data. Gray dots correspond to the upper bound $2\Delta J_{\unitb}^2/J$ extracted from the Fig.\,\ref{fig_variances}a data. 
(e) Measured metrological gain $G$ (blue dots) as a function of the interaction time $t$.
The gray diamonds correspond to the upper bound $2\Delta J_{\mathrm{max}}^2/J$ (from Fig.\,\ref{fig_variances}c), and the  red squares are the $\bar G$ values from Fig.\;\ref{fig_Ramsey}d. The solid blue and dahed red lines corresponds to the gains $G$ and $\bar G$ expected from the one-axis twisting model.
\label{fig_Hellinger}}
\end{figure}

To go beyond this `usual' metrological gain $\bar G$, we now exploit a key feature of our setup, i.e., the ability to resolve individual sublevels. This allows us to unveil small-scale structures in the measured projection probabilities $\Pi_m(\theta)$ that  rapidly vary with $\theta$, suggesting hidden phase sensitivity in higher-order moments of the probability distribution, even when $\bar G<1$. In order to quantify this $\theta$ dependence, we introduce the Hellinger distance between probability distributions $d_{\mathrm{H}}^2(\theta,\theta')\equiv\frac{1}{2}\sum_m[\sqrt{\Pi_m(\theta)}-\sqrt{\Pi_m(\theta')}]^2$. The phase sensitivity, expressed in terms of metrological gain,  is then related to the curvature of the Hellinger distance  as \cite{braunstein_statistical_1994-1,strobel_fisher_2014-1}
\begin{equation}\label{eq_GN}
G(\theta)=\frac{2}{J}\left.\frac{\partial^2 d_{\mathrm{H}}^2(\theta,\theta+\nu)}{\partial\nu^2}\right|_{\nu=0}.
\end{equation}
This gain coincides with the usual gain $\bar G$ for states with gaussian quantum fluctuations.

We show in Fig.\,\ref{fig_Hellinger}a,b  the projection probabilities $\Pi_m(\theta)$ measured for an  oversqueezed state (interaction time $t=0.84(1)\tau$). As theoretically shown in Ref.\,\cite{nolan_optimal_2017-1}, we expect, for this protocol, optimal sensitivity around $\theta=0$. We observe strong  population variations when the rotation axis $\unitb$ coincides with the direction $\unitn_{\mathrm{max}}$ of maximal spin projection variance (Fig.\,\ref{fig_Hellinger}b), and minor variations for $\unitb=\unitn_{\mathrm{min}}$ (Fig.\,\ref{fig_Hellinger}a).  To extract the metrological gain $G$, we calculate the Hellinger distances $d_{\mathrm{H}}^2(\theta,\theta')$ from the measured $\Pi_m(\theta)$ data and use a polynomial fit to extract its curvature around $\theta=\theta'=0$ \cite{Note1}.  We show in Fig.\,\ref{fig_Hellinger}c examples of cuts $d_{\mathrm{H}}^2(\theta,\theta'=0)$, together with the corresponding fits. As shown in Fig.\,\ref{fig_Hellinger}d, we find that the measured gain agrees well for all rotation axes $\unitb$ with the quantum Cram\'er-Rao bound for a pure  state -- the maximum achievable sensitivity -- given by $2\Delta J_{\unitb}^2/J$ \cite{braunstein_statistical_1994-1}. The optimal character of this measurement protocol has been demonstrated theoretically in Ref.\,\cite{nolan_optimal_2017-1}, and is based on the conservation of parity by the one-axis twisting Hamiltonian.
%

We repeat this measurement for  various interaction times up to $t=2\,\tau$   (see Fig.\,\ref{fig_Hellinger}e). For $t<0.5\,\tau$, the measured gain $G$ remains close to the usual gain $\bar G$ deduced from the first two moments, as expected in this regime of gaussian quantum fluctuations \cite{pezze_quantum_2014-1}. For longer times, the measured gain $G$ largely exceeds the gain $\bar G$, reaching an almost constant value $G=8.6(6)$ in the oversqueezed regime (average value of $t>\tau$ data). This value is consistent with $G=J+\frac{1}{2}$ expected for a spin state uniformly spanning the entire $yz$ meridian. The measured sensitivity closely follows  the one-axis twisting model prediction, and it remains close to the upper bound $(2/J)\Delta J_{\mathrm{max}}^2$ in the whole considered range of interaction times.

\begin{figure}
\hspace*{-0.02\linewidth}\includegraphics[
trim={1mm 0 0 2mm},
draft=false,
scale=0.8
]{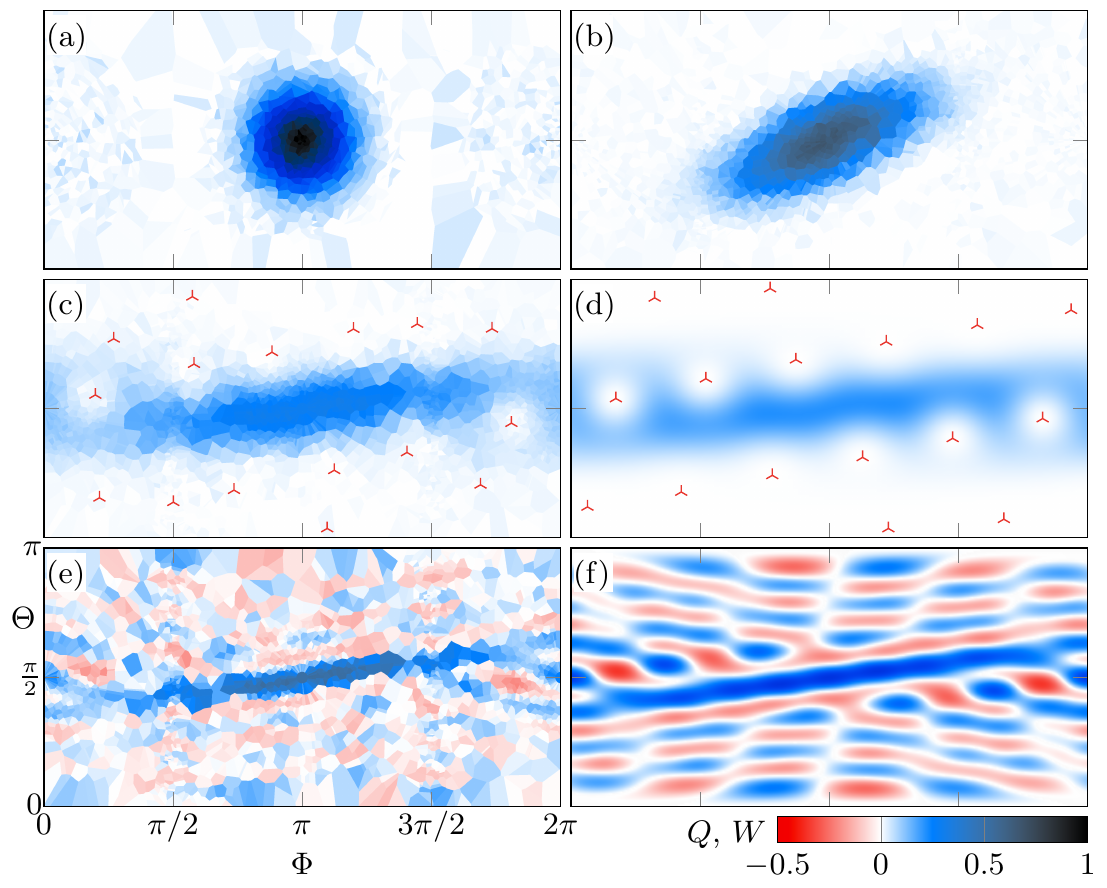}
\vspace{-7mm}
\caption{
(a,b,c) Husimi $Q$ function measured for a coherent, squeezed and oversqueezed spin states (a, b and c), achieved after evolution times $t/\tau=0$, 0.48(2) and 2.2(1), respectively.  The Bloch sphere is parameterized by the spherical angles  $(\Theta,\Phi)$ associated with the frame $(y,z,x)$. The red stars in (c) indicate the fitted zeros of the Husimi function. (d,f) Husimi (d) and Wigner (f) functions  of the quantum state expected from the one-axis twisting model for an interaction time $t=2.2\tau$.
(e) Wigner function reconstructed from the same data used in (c). 
\label{fig_Husimi_Wigner}}
\end{figure}

To get more physical insight we characterize the produced quantum states by their phase space representation on the generalized Bloch sphere. We consider in the following two quasi-probability distributions, the Wigner function $W$ and the Husimi function $Q$ \cite{wigner_quantum_1932,husimi_formal_1940}. The Wigner function, defined for a spin in \cite{dowling_wigner_1994}, is an indicator of non-classical behavior via its  negative-value regions. The Husimi function $Q(\unitn)$, defined as the squared overlap with a coherent spin state pointing along $\unitn$ \cite{husimi_formal_1940}, corresponds to a gaussian smoothening of the Wigner function \cite{hillery_distribution_1984}. We compute both functions from the measured probabilities $\Pi_m(\unitn)$, using $Q(\unitn)=\Pi_{m=J}(\unitn)$ and  $W(\unitn)=\sum_{m}(-1)^{J-m}a_m\,\Pi_{m}(\unitn)$,
with \mbox{$a_m\equiv\sum_{k=0}^{2J}(2k+1)\langle J,m,J,-m;k,0\rangle/\sqrt{4\pi}$} \cite{dowling_wigner_1994}.
As a reference, we measured the Husimi function of a coherent spin state (see Fig.\,\ref{fig_Husimi_Wigner}a). We find an almost isotropic gaussian distribution of r.m.s. angular width $\delta\theta=0.351(2)$, close to the expected value $1/\sqrt{J}\simeq0.354$. For  a short  time $t=0.48(2)\tau$, we reconstruct a twisted  Husimi function, well described by an anisotropic gaussian distribution (see Fig.\,\ref{fig_Husimi_Wigner}b).  For $t=2.2(1)\tau$, in the oversqueezed regime, the distribution has spread over the full $yz$ meridian (see Fig.\,\ref{fig_Husimi_Wigner}c). While semi-classical dynamics would predict diffusion towards a featureless distribution, we observe several small-scale dips that we interpret as the location of zeros of the Husimi function. For a pure quantum state $\ket\psi$ of a spin $J$, we expect the occurence of $2J$ zeros in the Husimi function, corresponding to the opposite orientations of the $2J$ fictitious spin-$1/2$ particles composing the spin $J$ -- the so-called Majorana stellar representation \cite{majorana_atomi_1932}. Denoting these orientations $\unitu_i$ ($1\leq i\leq 2J$), the Husimi function reads 
$
Q(\unitn)\propto\prod_i(1+\unitu_i\cdot\unitn),
$
 and vanishes for $\unitn=-\unitu_i$ \cite{Note4}. Fitting the entire distribution with this ansatz, we obtain the location of all zeros of the Husimi function, in good agreement with the expected positions (see Fig.\,\ref{fig_Husimi_Wigner}d). 
We show in Fig.\,\ref{fig_Husimi_Wigner}e the Wigner function reconstructed for the oversqueezed state. It exhibits negative values in a large fraction of phase space, indicating a highly non-classical character \cite{kenfack_negativity_2004-1}. We also find small-scale oscillations reminiscent of `sub-Planck' structuring of phase space, as expected for metrologically useful quantum states  \cite{zurek_sub-planck_2001}. While the measured small-scale structures in the Husimi function are not directly linked to the magnetic  sensitivity, the oscillations found in the Wigner function imply a fast variation of the state upon rotation, making a direct connection with the high magnetic sensitivity of oversqueezed states \cite{paris_quantum_2009}.

To conclude, we showed that measurements based on single magnetic sublevel resolution allow reaching optimal sensitivity with non-gaussian states of a quantum spin $J$. An optimum $G=8.6(6)$ is reached  as soon as the spin distribution is stretched along the full $yz$ meridian. The Heisenberg limit $G=16$ could in principle be achieved using the maximally entangled N00N state \cite{monz_14-qubit_2011-1,chalopin_quantum-enhanced_2018}; however, the required interaction time $t=\sqrt{\pi^2J/2}\,\tau$ is much longer than $\tau$ for $J\gg 1$, making this state more fragile  to decoherence \cite{Note1}. Oversqueezed states thus appear as a compromise for future progress with large atomic ensembles. 
We also provided a full characterization of non-classical spin states in phase space in terms of their Majorana stellar representation. The latter could be used to characterize ordering in spinor quantum gases \cite{stamper-kurn_spinor_2013-2}, geometric quantum entanglement \cite{liu_representation_2014-1} or chaotic behavior \cite{leboeuf_chaos-revealing_1990}. 

\begin{acknowledgments}
We thank Chayma Bouazza for contributions in earlier stages of the experiment. This work is supported by PSL University (MAFAG project) and European Union (ERC
UQUAM and TOPODY, Marie Curie project 661433). 
\end{acknowledgments}

%

\footnotetext[1]{See Supplemental Material  for details on the metrological gain in presence of noise, experiment protocols, additional Hellinger distance data, Husimi and Wigner functions measured and computed for coherent and gaussian squeezed states, and a discussion on the evaluation of purity of quantum states, which includes
Refs. \cite{dariano_spin_2003-1,filippov_purity_2013}.}

\footnotetext[3]{We estimate relative corrections to this coupling due to imperfect light polarization to remain below 0.05\%. Furthermore, the laser detuning  $\Delta=-2\pi\times\SI{1.1(1)}{GHz}$ ensures negligible incoherent Raman scattering over the typical light pulse duration $t\sim\SI{100}{\nano\second}$.}

\footnotetext[4]{We also checked that the spin state remains quasipure on this timescale \cite{Note1}.}

\end{document}


\title{Supplemental Material for: \\ Enhanced magnetic sensitivity with non-gaussian quantum fluctuations 
}

\author{Alexandre Evrard}
\author{Vasiliy Makhalov}
\author{Thomas Chalopin}
\author{Leonid A. Sidorenkov}
\altaffiliation{LNE-SYRTE, Observatoire de Paris, Universit\'e PSL, CNRS, Sorbonne Universit\'e, 61 Avenue de l'Observatoire, 75014 Paris, France}
\author{Jean Dalibard}
\author{Raphael Lopes}
\author{Sylvain Nascimbene}
\email{sylvain.nascimbene@lkb.ens.fr}

\affiliation{Laboratoire Kastler Brossel,  Coll\`ege de France, CNRS, ENS-PSL University, Sorbonne Universit\'e, 11 Place Marcelin Berthelot, 75005 Paris, France}

\maketitle 

\section{Robustness over non-linear coupling fluctuations}

In the main text we discussed the realization of  oversqueezed states for interaction times $t\sim\tau$, leading to a metrological gain $G\simeq8.5$. The Heisenberg limit  $G=16$ can in principle be reached at a time $t = \pi \sqrt{J/2} \tau$, corresponding to a Schr\"odinger cat state. We discuss here the robustness of the metrological gain for these two states with respect to fluctuations of the non-linear coupling $\chi$. 

\begin{figure}[h!]
	\hspace*{-0.05\linewidth}\includegraphics[width=1\linewidth]{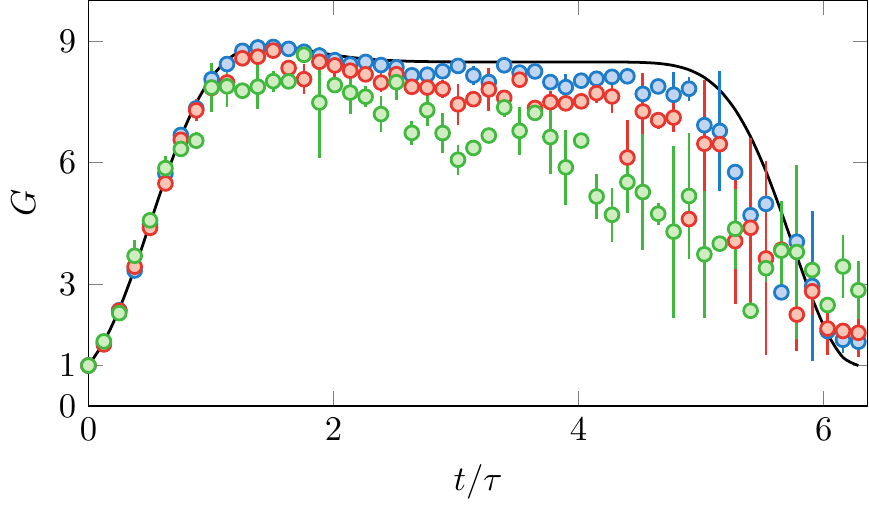}\\
	\caption{\label{Fig1_SI}Metrological gain as a function of the interaction time $t$, for non-linear coupling fluctuations of $0$ (black line), $5\%$ (blue marks), $10\%$ (red marks) and 20$\%$ (green marks).}
\end{figure}

We show in Fig.~\ref{Fig1_SI} the metrological gain $G$, numerically extracted following the experimental procedure discussed in the main text, as a function of $t$, in the absence of noise (black line). We then compute the gain $G$ averaged over shot-to-shot fluctuations of $\chi$ with different r.m.s. amplitudes: 5$\%$ (blue), 10$\%$ (red), and $20 \%$ (green). We observe that even in the case of large fluctuations (20 $\%$) the expected metrological gain remains close to its ideal value at $t=\tau$, falling solely to 7.8.

For the cat state a different procedure is required, as explained in Ref.~\cite{chalopin_quantum-enhanced_2018}. In that case, 
for fluctuations' amplitudes of $\chi $ = $20 \%$, the metrological gain falls from an ideal value of 16 to a value of approximately 9.

\section{Experimental protocol}
	
\textbf{Imaging calibration.} Our imaging setup is such that the 17 magnetic sublevels have different cross-sections. To calibrate their effective Clebsh-Gordan coefficients, we prepare a set of spin-polarized samples with constant total atom number and various orientations on the Bloch sphere. We then choose the effective Clebsh-Gordan coefficients  to minimize the variations of the calculated total number of atoms over the different state orientations. 

\textbf{Calibration of the diffusion time} $\mathbf{\tau}$\textbf{.} We calibrate the diffusion time $\tau$ by applying the non-linear coupling for a set of interaction times $t$ and measuring the projection probabilities $\Pi_m(\hat{\mathbf{z}})$. We fit the measured probabilities with the variations expected from the one-axis twisting model, the diffusion time $\tau$ being the only free parameter. We checked that the fitted value of $\tau$ is consistent with a direct calculation based on the measured laser intensity and frequency detuning from the atomic resonance.

\textbf{Magnetic pulses.} Arbitrary spin rotations are performed by combining the free spin precession around the quantization field along $z$, of amplitude $B=\SI{60.6(3)}{mG}$, with `gate' pulses of magnetic field along $y$, applied for a duration $\simeq\SI{3}{\micro s}$. We mention that during the pulse duration the $z$ magnetic field induces a non-negligible Larmor rotation, equal to \SI{1.94}{rad}, that we take into account in the determination of the rotation angles.

\textbf{Correction of magnetic field fluctuations.} We first use an active stabilization of the magnetic field to decrease residual magnetic field fluctuations below \SI{0.5}{mG} (r.m.s. value). The remaining fluctuations are measured using a three-axis magnetic field probe placed about 10 cm away from the atoms. They evolve on a timescale of $\gtrsim\SI{10}{ms}$, slow compared to the entire spin evolution, and can therefore be treated as a constant.  We account for this offset in the data analysis for the calculation of Larmor rotation angles.

\section{Fit of the Hellinger distance data} The metrological gain $G$ is extracted from the measured Hellinger distance in the following way. We fit the experimental data with a polynomial in the two variables $(\theta + \theta')^2$ and $(\theta - \theta')^2$ of order $n$, as

\begin{equation}
d^2_{\mathrm{H}}(\theta, \theta') = \sum_{0\leq p+q \leq n} a_{p,q}(\theta - \theta')^{2p}(\theta + \theta')^{2q}.
\end{equation}

\begin{figure}
	\hspace*{-0.05\linewidth}\includegraphics[width=1\linewidth]{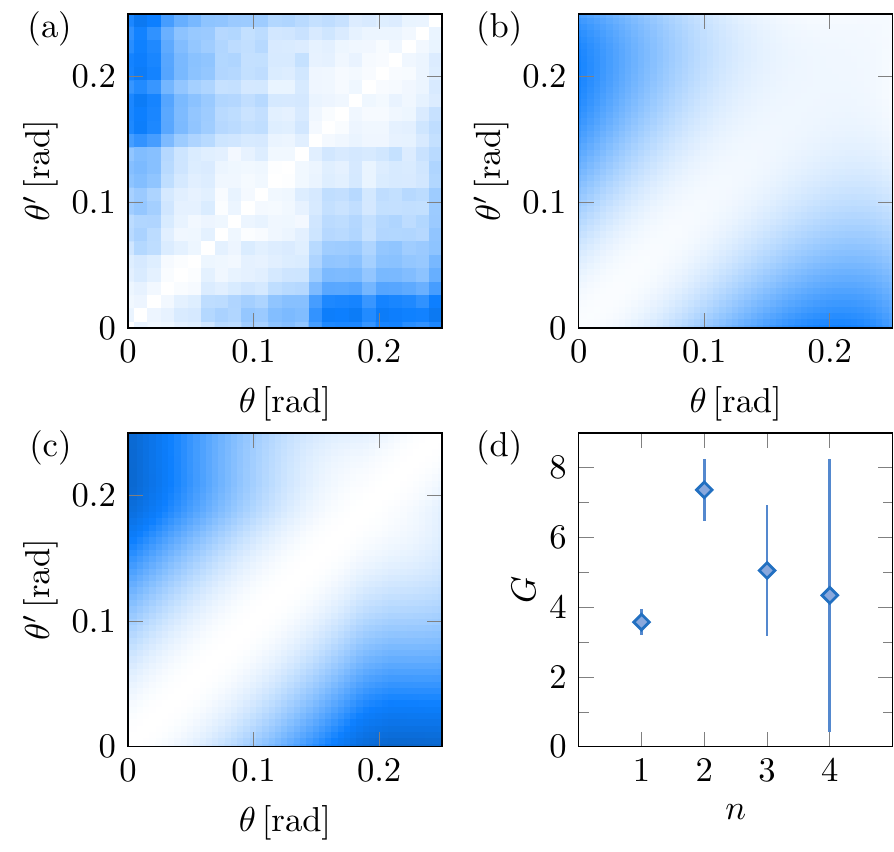}\\
	\caption{\label{Fig2_SI} (a) Hellinger distance measured for an interaction time $t=0.835\,\tau$. (b) Fit of the measured distances with a polynom of order $n=2$. (c) Hellinger distance expected from the one-axis twisting model for the same interaction time. (d) Metrological gain $G$ deduced from (b) as a function of the order of the fitting polynomial.}
\end{figure}

The metrological gain $G$ is deduced from the curvature of the fit around $\theta=\theta'=0$, as $G=4a_{1,0}/J$ (see Eq. (4) of the main article). We tested several values of the order $n$. As shown  in Fig.~\ref{Fig2_SI}d, an order $n=1$  produces significant systematic shifts, while orders $n\geq3$ leads to an increase of statistical error bars. We thus choose the value $n=2$ for all data analysis presented in the main article. An example of  Hellinger distance data, corresponding to the data shown in Fig.~4b of the main article, is displayed in Fig.~\ref{Fig2_SI}a, together with its polynomial fit (Fig.~\ref{Fig2_SI}b) and the values predicted within the one-axis twisting model for $t=0.835\,\tau$ (Fig.~\ref{Fig2_SI}c).

\section{Husimi and Wigner functions} We provide here both the theoretical and measured Husimi functions along with their fitted zeroes (red stars) and Wigner functions for a coherent state and a squeezed state, corresponding to interaction times $t=0$ and $t=0.48(2)$, respectively.

\begin{figure}
	\hspace*{-0.02\linewidth}\includegraphics[
	trim={1mm 0 0 2mm},
	draft=false,
	scale=0.8
	]{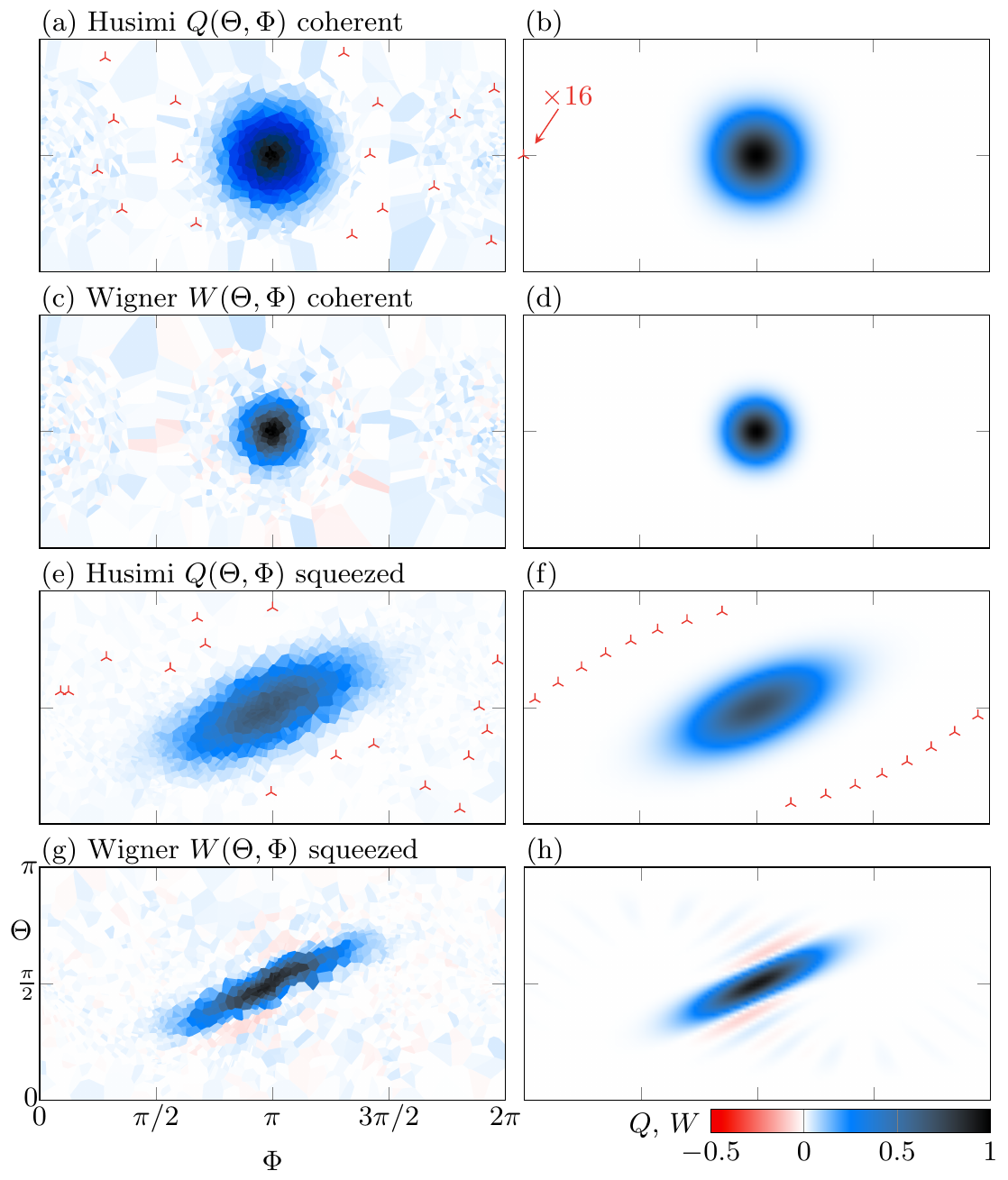}
	\caption{\label{Fig3_SI}Measured (left) and theoretical (right) Husimi and Wigner functions for coherent and squeezed states (interaction times $t=0$ and $t=0.48(2)\,\tau$, respectively).}
\end{figure}

The zeroes are not fitted correctly for the coherent state (see Fig.~\ref{Fig3_SI}a), for which we expect all zeros to be located at the north pole (see Fig.~\ref{Fig3_SI}b). However, the quantum state corresponding to the fitted Husimi function has a squared overlap of 0.99 with the coherent state $\ket{-J}_z$. This illustrates the difficulty to determine the position of the zeroes of the Husimi function located in small-amplitude regions due to background noise. The zeroes are better fitted for a squeezed state, as they are located closer to large-amplitude regions of phase space  (see Fig.~\ref{Fig3_SI}e,~\ref{Fig3_SI}f).

We also show the Wigner functions for a coherent and squeezed state in Fig.~\ref{Fig3_SI}c,~\ref{Fig3_SI}g. For the squeezed state, we measure faint fringes with negative-amplitude regions consistent with theory. 

\section{Purity measurement} The measured spin projection probabilities $\Pi_{m}(\hat{\mathbf{n}})$ can be used to directly evaluate the purity of the prepared spin states without having to reconstruct the full density matrix $\rho$.

\begin{figure}
	\hspace*{-0.0\linewidth}\includegraphics[width=1\linewidth]{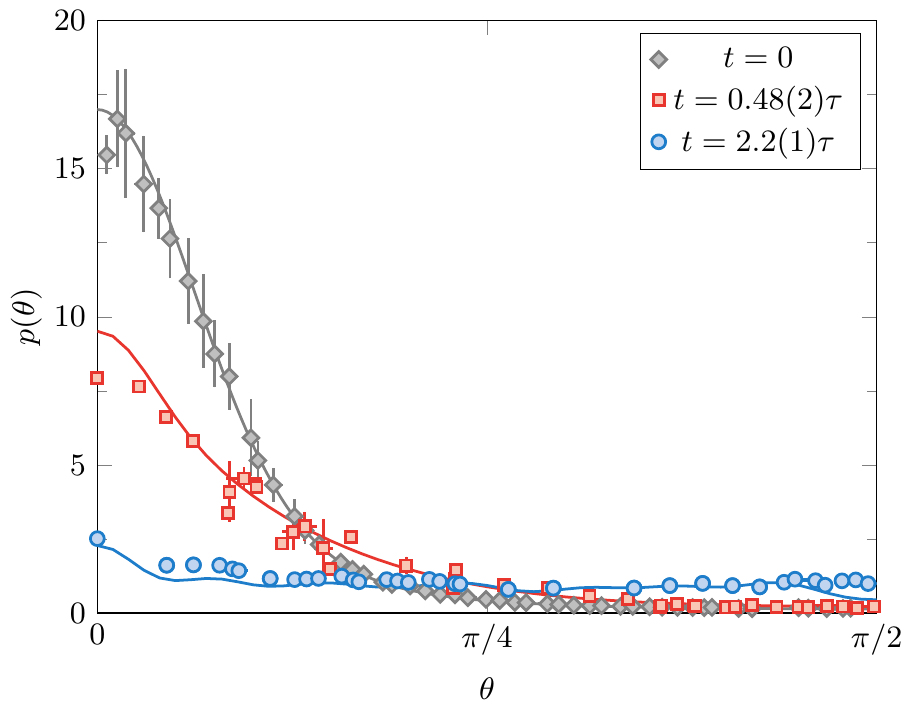}\\
	\caption{\label{Fig4_SI}Values of $p(\theta)$ computed from the $\Pi_m(\theta,\phi)$ data for a coherent, squeezed and oversqueezed state (gray, red and blue symbols,  interaction times $t=0$, $t=0.48(2)\,\tau$, and $t=2.2(1)\,\tau$ respectively). The solid lines correspond to the prediction of the one-axis twisting model.}
\end{figure}

According to spin tomography theory, the density matrix of a quantum spin can be reconstructed from integration of the probabilities $\Pi_{m}(\hat{\mathbf{n}})$ over the Bloch sphere, as \cite{dariano_spin_2003-1}
\begin{align}
\rho&=\sum_{m=-J}^J\int_{S^2} \frac{\text{d}\hat{\mathbf{n}}}{4\pi} \, \Pi_{m}(\hat{\mathbf{n}}) \text{D}(m, \hat{\mathbf{n}}),\\
 \text{D}(m, \hat{\mathbf{n}})&=\frac{2J+1}{\pi}\int_{0}^{2\pi}\text{d}\gamma \, \sin^2\frac{\gamma}{2} \, \text{e}^{i\gamma(m-\hat{\mathbf{J}}\cdot\hat{\mathbf{n}})},
\end{align}
where $S^2$ denotes the unit sphere.
The purity can then be written as \cite{filippov_purity_2013}
\begin{align}
&\text{Tr}\!\left[\rho^2\right]=\sum_{m=-J}^J\int_{S^2} \frac{\text{d}\hat{\mathbf{n}}}{4\pi} \, \Pi_{m}(\hat{\mathbf{n}})\, \text{Tr}\left[\rho \, \text{D}(m, \hat{\mathbf{n}})\right]\\
&=(2J+1)\sum_{m=-J}^J\int_{S^2} \frac{\text{d}\hat{\mathbf{n}}}{4\pi} \, \Pi_{m}(\hat{\mathbf{n}}) \left[\Pi_{m}(\hat{\mathbf{n}})-\Pi_{m+1}(\hat{\mathbf{n}})\right],\label{eq_purity}
	\end{align}
	where we define $\Pi_{J+1}(\hat{\mathbf{n}})=0$.
We evaluate the integral (\ref{eq_purity}) from a discrete set of $\simeq 1000$ independent $\Pi_{m}(\hat{\mathbf{n}})$ measurements sampling the sphere $S^2$. We show in Fig.~\ref{Fig4_SI} the quantity $p(\theta)$  obtained after a first integration over the azimutal angle $\phi$, i.e. 
\begin{equation}
p(\theta)=(2J+1)\sum_{m=-J}^J\int_{0}^{2\pi} \frac{\text{d}\phi}{2\pi} \, \Pi_{m}(\hat{\mathbf{n}}) \left[\Pi_{m}(\hat{\mathbf{n}})-\Pi_{m+1}(\hat{\mathbf{n}})\right],
\end{equation}
such that
\begin{equation}
\text{Tr}\!\left[\rho^2\right]=\frac{1}{2}\int_0^{\pi} p(\theta)\sin\theta\,\text{d}\theta.
\end{equation}
The quantity $p(\theta)$ measured for coherent, squeezed and oversqueezed states matches well the expected values within the one-axis twisting model. Integrating over the variable $\theta$, we obtain purity values $\text{Tr}\!\left[\rho^2\right]=1.00(2)$, 1.00(3) and  1.01(4) respectively. Thanks to the symmetry $\Pi_m(\pi,\phi)=\Pi_{-m}(\pi-\theta,\phi+\pi)$, it is sufficient to compute $p(\theta)$ for $\theta < \pi/2$. The error bars are determined using a bootstrap sampling method.